\documentclass[12pt]{article}
\usepackage{amsmath}
\usepackage{amsfonts}
\usepackage{amscd}
\usepackage{eucal}

\newcommand{\ddouble}{{\partial^{^{\kern-6pt \leftrightarrow}}}}

\newcommand{\es}{\\[3mm]}
\newcommand{\beq}{\begin{equation}}
\newcommand{\eqn}[1]{\label{#1}\end{equation}}
\newcommand{\ba}{\begin{array}}
\newcommand{\ea}{\end{array}}

\def\psi{\Psi}

\def\inbar{\vrule height1.5ex width.4pt depth0pt}
\def\rlx{\relax\leavevmode}
\def\I{\leavevmode\hbox{\small1\kern-3.8pt\normalsize1}}
\def\openone{\leavevmode\hbox{\small1\kern-3.3pt\normalsize1}}
\def\Ione{\rlx{\rm 1\kern-2.7pt l}}
\def\Ik{\rlx{\rm I\kern-.18em k}}
\def\IC{\rlx\leavevmode
             \ifmmode\mathchoice
                    {\hbox{\kern.33em\inbar\kern-.3em{\rm C}}}
                    {\hbox{\kern.33em\inbar\kern-.3em{\rm C}}}
                    {\hbox{\kern.28em\sinbar\kern-.25em{\rm C}}}
                    {\hbox{\kern.25em\ssinbar\kern-.22em{\rm C}}}
             \else{\hbox{\kern.3em\inbar\kern-.3em{\rm C}}}\fi}
\def\IP{\rlx{\rm I\kern-.18em P}}
\def\IR{\rlx{\rm I\kern-.18em R}}
\def\IN{\rlx{\rm I\kern-.20em N}}

\def\llsymbol#1{\@llsymbol{\@nameuse{c@#1}}}
\def\@llsymbol#1{\ifcase#1\or {}\or {'}\or {''}\or {'''}\or
   {''''}\or {'''''}\or  \else\@ctrerr\fi\relax}
\newcounter{contador}

\newcommand{\ol}\overline
\newcommand{\ti}\tilde
\newcommand{\wt}\widetilde
\newcommand{\wh}\widehat
\newcommand{\bv}\breve
\newcommand{\dg}\dagger

\newcommand{\C}{^{\mbox{\scriptsize c}}}

\newcommand{\be}{\begin{equation}}
\newcommand{\ee}{\end{equation}}
\newcommand{\bl}{\begin{eqnarray}&}
\newcommand{\el}{&\end{eqnarray}}
\newcommand{\bq}{\begin{eqnarray}}
\newcommand{\eq}{\end{eqnarray}}

\def\ptoday{{\ifcase\month 
\or January, \or February, \or March, \or April,\or May, 
\or June, \or July, \or August, \or September, \or October, 
\or November, \or December,\fi\ \number \year}}

\begin{document}

{\hfill
\parbox{40mm}{{ICEN-PS-00/00} \\ {\ptoday} \vspace{4mm}}

\begin{center}
{{\LARGE {\rm Asymptotic Conformal Invariance in a Non-Abelian 
Chern-Simons-Matter Model}}} 
\vspace{5mm}

{\large J.L. Acebal} 

\vspace{0,5cm}

{\em Centro Brasileiro de Pesquisas F\'\i sicas - Coordena\c c\~ao de Campos
 e Part\'\i culas} 

\vspace{0,5cm} 

{\tt e-mail: acebal@cbpf.br}

\end{center}

\begin{abstract}
One shows here the existence of solutions to the Callan-Symanzik equation for
the non-Abelian $SU(2)$ Chern-Simons-matter model which exhibits asymptotic 
conformal invariance to every order in perturbative theory. The conformal
symmetry in the classical domain is shown to hold by means of a local criteria 
based on the trace of the energy-momentum tensor. By using the recently 
exhibited regimes
for the dependence between the several couplings in which the set of 
$\beta$-functions vanish, the asymptotic conformal invariance of the model 
appears to be valid in the quantum domain. By considering  the $SU(n)$ case 
the possible non validity of the proof for a particular $n$ would be 
merely accidental.  
\end{abstract}

\vspace{0,3cm}

\small{

keywords: Asymptotic Conformal Invariance, Renormalization, Chern-Simons-matter} 

Topological field theories\footnote{See~\cite{bbr-pr91} for a general 
review and references.} are a class of gauge models which has deserved 
much attention, for they have an interesting ultraviolet behavior
that renders easier the task of getting non-perturbative results.  
In particular, the Chern-Simons model in three dimensions~\cite{s-btc} has been
deeply studied in the lastest years. They are sometimes presented as possible
applications in super-conductivity and $3D$-gravity models~\cite{witten-grav}.
The relevant property of this model lies on a very interesting perturbative 
behavior, namely, its ultraviolet finiteness~\cite{witten-grav,guadagnini}. 
This is rigorously proven for 
the $D=1+2$ Chern-Simons theories in the Landau gauge~\cite{blasi} and for 
the Chern-Simons-Yang-Mills theory in~\cite{CFNP0}. The Abelian Chern-Simons
gauge field coupled to scalar matter field is shown to have trivial 
$\beta$-function in~\cite{maggiore}. In~\cite{CFNP}, from the algebraic
renormalization 
approach~\cite{pisor}, the non-Abelian theory coupled to both scalar and
spinorial matter fields is checked to have the Chern-Simons coupling 
constant kept unrenomalizable. The scale invariance of non-Abelian 
Chern-Simons minimally coupled to matter has already been discussed 
in the two-loop approximation in ref.~\cite{CSW}.
Recently, in ref.~\cite{zeda}, the non-Abelian-Chern-Simons gauge model
minimally coupled to matter, with matter-matter interaction 
terms compatible with the power-counting renormalizability criterion
is shown to exhibit asymptotic scale invariant solutions to the
Callan-Symanzik equation. A discussion on the asymptotic conformal
invariance is presented in \cite{Odin}.

In this work, the $SU(2)$ Chern-Simons-matter model minimally coupled to matter,
with matter-matter interaction terms compatible to the power-counting 
renormalizability criterion is considered. Dimensionful
couplings are also allowed. The existence of asymptotic conformal invariant
solutions to the Callan-Symanzik for the model is ascertained.
Conformal symmetry is obtained in
the classical domain by means of a suitable criterion \cite{Polc} based in a 
local trace relation of the energy-momentum tensor. Such a criterion simplifies
the task of extending the proof of the symmetry to the quantum domain. 
This extension is obtained under the light of the algebraic renormalization 
approach~\cite{pisor} and based in the results of the paper  \cite{CFNP}. 
Finally, from the recently obtained result, \cite{zeda}, namely, the existence
of solutions to the Callan-Symanzik equation
for which the several $\beta$-functions vanish, one proves the existence of 
conformal asymptotic invariant solutions to the Callan-Symanzik equation.      

The Chern-Simons-matter BRS covariant model in the Landau gauge 
has the following action, 
\begin{eqnarray}
\lefteqn{\Sigma\, = \int d^3 x \,\,\left\{ \kappa \,\varepsilon^{\mu \nu \rho}
      (A_\mu ^a\partial_\nu A_\rho ^a+\frac 13f_{abc}A_\mu^a A_\nu^b A_\rho^c)
	  \right.  
 + (i\,e\,\bar{\Psi}_{j}\gamma^\mu {\cal D}_\mu \Psi_j 
     -m_\Psi\bar{\Psi}_{j}\Psi_{j})+\nonumber}\\ 
&+&\mbox{\hspace*{-.3cm}}
\frac 12 (e\,{\cal D}_\mu \varphi_i^* {\cal D}^\mu \varphi_i 
- m_\varphi\varphi_j^*\varphi_{j})
 - \frac 12 \lambda_1 \bar{\Psi}_{j}\Psi_{j}\varphi_{k}^*\varphi_{k}
-\frac 12 \lambda_2 \bar{\Psi}_{j}\Psi_{k}\varphi_{k}^*\varphi_{j}  
 + \frac 16 \lambda_3(\varphi_i^*\varphi_{i})^3 \label{g-inv-action1}\\[2mm]
&+&\mbox{\hspace*{-.15cm}} (\mbox{``dim'ful'' couplings})+
e\,g^{\mu\nu} (\partial_\mu b_a A_\nu^a + \partial_\mu \bar{c}_a 
{\cal D}_\nu c^a)  
\,\,
+\mbox{\hspace{-2 em}} \sum_{\,\,\,\Phi=A_\mu ^a,\,c^a,\,\Psi_j,\,\varphi_i}
\left.\mbox{\hspace{-2 em}}\Phi^{\natural} s\Phi \,\right\}\,. 
\nonumber
\end{eqnarray}
The $e$ denotes the determinant of the dreinbein $e^m_\mu$.
The gauge field $A_\mu^a(x)$ lies in the adjoint representation of the gauge 
group $SU(n)$, with Lie algebra $\left[ X_a,X_b\right] =if_{ab}{}^cX_c$. 
The scalar matter fields, $\varphi _i(x)$, and the spinor matter fields, 
$\Psi_j(x)$, are in the fundamental representation of $SU(n)$. The 
generators are respectively the matrices $T_a^{(\varphi )}$ and 
$T_a^{(\Psi)}$. The fields $c^a$, $\bar{c}^a$ and $b^a$ are the ghost, 
the antighost and the Lagrange multiplier fields. The ${A}_{a\mu}^\natural$,
${c}_a^\natural$, ${\Psi}_j^\natural$, 
${\varphi}_i^\natural$ are the ``antifields'' coupled to the nonlinear 
variations under BRS transformations. The diffeomorphic form of the action
is considered in order to allow the determination of the Belifant tensor.
Indeed, the result is obtained in flat space-time limit.
The generalized covariant derivative is defined by 
\begin{equation}
{\cal D}_\mu \Phi(x)=( \partial _\mu -i\,{A_\mu^a}(x)T_a^{(\Phi )}+
\frac{1}{2}\omega^{mn}_{\mu}\Omega_{mn})\Phi (x)
\,\,. 
\end{equation}
The vanishing torsion makes the spin connection, $\omega^{mn}_{\mu}(x)$, 
dependent on the dreibein $e^m_\mu(x)$. One considers a manifold
with asymptotically flat curvature and topology equivalent to the 
flat ${\cal R}^3$.            
The various symmetries of the model, manifested through its Ward 
identities are BRS, local Lorentz transformations, diffeomorphism and rigid
gauge invariance, besides the constraints, Landau gauge condition, ghost 
and antighost equations. They are all shown in the work of 
ref.~\cite{CFNP}. 

The symmetric energy-momentum tensor,
$\Theta^\mu_\nu$,
can be defined as  
\begin{equation}
{\Theta }_{\mu\nu} = e^{-1} e_{(\mu}^{~m}~\frac {\delta
\Sigma}{\delta e^{\nu)m}}\,\,.  
\label{theta}
\end{equation}
It is well-known that every space-time symmetry may be related to the 
energy-momentum tensor. In particular, the trace $\Theta^\mu_\mu$ is deeply
connected 
to both dilatation and conformal symmetries \cite{Polc}. By integrating, one
obtains the Ward identity for the dilation invariance,
\begin{eqnarray}
& & \int\,d^3x\,e\Theta^\mu_\mu=\int\,d^3x\, e_\mu ^{~m}~
\frac {\delta \Sigma}{\delta e_\mu ^{~m}}={\cal N}_e \Sigma\sim\nonumber \\
& & \sim\underbrace{({\cal N}_\psi +{\cal N}_{\bar{\Psi}}
+\frac 12{\cal N}_{\varphi} + {\cal N}_b + {\cal N}_{\bar{c}}
-{\cal N}_{\Psi^{\natural}} - {\cal N}_{{\bar{\Psi}}^{\natural}}
-\frac 12{\cal N}_{\varphi ^{\natural}})\Sigma}_{
\int\,d^3x\,w^{(\Phi)}(x)\Sigma},
\label{int-theta1} 
\end{eqnarray}
where the definition of the counting operator, ${\cal N}_\Phi \Sigma
=\int\,d^3x\, \Phi~
\frac {\delta \Sigma}{\delta \Phi}$ with $(\Phi=\mbox{any field})$ is used
and  ``$\sim$''  means equality up to mass terms and dimensionful couplings.
The difference between the integrands of the expression above must be of the
form    
\begin{equation}
e \Theta_\mu{}^\mu(x) \sim w^{(\Phi)}(x)\Sigma 
+ \partial_\mu\Lambda^\mu(x)\,\,, 
\label{cl_trace}
\end{equation}
where, from \cite{CFNP},
\begin{equation}
\Lambda^\mu = e\,i\bar{\Psi}\gamma ^\mu \Psi 
+e \varphi\nabla^\mu \varphi -s\,\left( e\,g^{\mu \nu }\bar{c}A_\nu \right) 
\,\,. 
\label{a17}
\end{equation}

In what concerns the establishment of the conformal invariance in the classical
domain, one can not ignore the local character of the symmetry. There are 
interesting criteria \cite{Zumi,Iorio} based on the symmetries manifested in the
diffeomophic version of the theory, but somewhat difficult to be interpreted at
the quantum level.
In the previous analysis~\cite{CFNP,zeda} of the model (\ref{g-inv-action1}), 
the local trace identities, related to the Callan-Symanzik equation take a
relevant place. Hence, it would be suitable to use a criterion~\cite{Polc}
which points in such a direction. In focusing the conservation of 
energy-momentum, $\partial^\mu T^{\prime}_{\mu\nu}=0$, and definition of the
charge $P_\mu=\int d^{d-1}x\,T_{0\mu}$, the tensor can admit different 
definitions \cite{Polc, CaCoJa, Gatto}. 
In considering space-time transformations in the form of general
coordinate transformations (GCT) $x^{\prime\mu}=x^{\mu}+
\varepsilon^{\mu}(x)$, the complete ten-parameters $(1+2)D$ conformal 
transformations can be mapped into GCT by the relation~\cite{ZJ},
\begin{equation}
\partial_{\mu}\varepsilon_{\nu} + \partial_{\nu}\varepsilon_{\mu} =
\frac{2}{d}\eta_{\mu\nu}(\partial\cdot\varepsilon)\mbox{\hspace*{0.3cm}}
\Longrightarrow \mbox{\hspace*{0.3cm}}\partial^2(\partial\cdot\varepsilon)=0\,.
\label{15conf}
\end{equation}
The second order solutions are the special conformal group. The first
order are the Poincar\'{e} group and the zeroth order are the dilatations.
The current associated to the complete conformal group can be written as
\begin{equation}
j^\mu=\varepsilon^\nu(x)T^\mu_\nu(x)+(\partial\cdot\varepsilon)K^\mu
+ \partial_\nu(\partial\cdot\varepsilon)L^{\mu\nu} +
(\partial_\nu\varepsilon_\rho-\partial_\rho\varepsilon_\nu)M^{[\nu\rho]\mu},
\label{gen_curr}
\end{equation}
where $K$,$L$ and $M$ local quantities depending on the fields and its
first derivatives. The usual forms \cite{CaCoJa, Gatto} can be obtained by 
imposing the various forms of $\varepsilon_\nu(x)$, with respect to the  
subgroups of the whole conformal group. In \cite{Polc}, it is shown that in 
considering the dilatations $\varepsilon_\nu(x)=\lambda x_\nu$, under the 
requirement $\partial_\mu j^\mu=0$, there arises the condition 
$T^{~\mu}_{\mu}=-\partial_\mu K^\mu$, and that, for the restricted
conformal  group $\varepsilon_\nu(x)=a_ \nu x^2-2(a\cdot x)x_\nu$,
the requirement $\partial_\mu j^\mu=0$ furnishes both 
$T^{~\mu}_{\mu}=-\partial_\mu K^\mu$ and $K^\mu=-\partial_\mu L^{\mu\nu}$. 
Hence, the condition for conformal invariance is shown to be stronger than
the condition for scale invariance. The conditions for conformal invariance
taken together yield the sufficient condition
\begin{equation}
T^\mu_\mu=\partial_\mu \partial_\sigma L^{\mu\sigma}.
\label{ptrace}
\end{equation}
If the requirements above are satisfyied, there follows the Ward identity 
for the conformal group. Redefinitions on $T^{\mu\nu}$ imply similar 
redefinitions on $K^\mu$ and $L^{\sigma\rho}$. 
The traceless $T^{\mu\nu}$ is a particular case
obtained for the improved tensor. The existence of a traceless
$T^{\mu\nu}$ depends on the theory and is a particular case of the present
criterion. Then, it can be stated, \cite{Polc}, that the existence of a 
traceless stress tensor is equivalent to the conformal invariance.  

From now on, the study will be restricted to the flat space limit. 
By focusing the attention on the r.h.s of the
expression (\ref{cl_trace}) and taking the flat limit space, one notes that
the first term is a set of equations of motion which vanish on-shell. 
The second term in (\ref{cl_trace}) can be studied in the expression 
(\ref{a17}). In (\ref{a17}), the third term is a total BRS variation 
and is again physically trivial, being non-observable on-shell. Hence, the trace 
(\ref{cl_trace}), up to terms which vanish on-shell is given by
\begin{equation}
\Theta^\mu_\mu\sim i\partial_\mu(\bar{\Psi}\gamma^\mu\Psi) + 
\frac{1}{2}\partial_\mu \partial_\sigma \eta^{\mu\sigma} (\varphi^2).
\end{equation}
By invoking the global gauge symmetry both $\bar{\Psi}\gamma^\mu\Psi$ 
and $\varphi^2$ show to be conserved quantities, what causes the 
the vanishing of the trace of the energy-momentum tensor 
$\Theta^\mu_\mu\sim 0$ up to mass terms
and dimensionful couplings. Therefore, the
condition for the asymptotic on-shell conformal invariance holds in the
classical domain.

The possible validity of the conformal invariance for (\ref{g-inv-action1}) 
into the quantum domain would depend on the existence, or not, of anomalies. 
This matter was examined by~\cite{CFNP} in
the framework of the algebraic renormalization method~\cite{pisor}. 
In such an approach, the study of the renormalizability was carried out through
the analysis of the extension of the Slanov-Taylor (ST) identity to the quantum
domain.  The absence
of gauge anomalies for diffeomorphisms and local Lorentz invariant functionals
in this class of manifolds, \cite{Bran,Bar}, in addition to the
power-counting renormalizability criterion, restrict that analysis to the 
space of the diffeomorphism and local Lorentz invariant integral functionals
$\Delta$ of the various fields having integrand of dimension 3. 
The gauge condition, ghost, antighost equations and rigid gauge invariance
are shown to be non-anomalous~\cite{pisor}.
The extension of the validity of
the various symmetries to the vertex functional, $\Gamma$, is restricted to
the analysis of the basis generated by the problem posed by the ST
identity, $B_{\Sigma}\Delta=0$, $B_{\Sigma}$ being the ST operator, 
with conditions imposed by the diffeomorphism, local Lorentz
and rigid gauge symmetries, together with the constraints given by the Landau
gauge condition, ghost and antighost equations. 
The solution of this given problem generates then a basis in functional integral
space where various studies, renormalizability, non-renormalization of
the parameters of the theory, anomalies could take place. 
The solutions
are divided in two parts: 
$\Delta=\Delta_{\rm cohom} + B_{\Sigma}\hat{\Delta}$. 
Because the nilpotency, $B^2_{\Sigma}=0$, the second term is the trivial part,  
and is associated to non-physical field redefinitions. The 
$\Delta_{\rm cohom}$ part is associated to the renormalization of physical 
parameters. The space of the solutions are also divided in sectors, depending
on the ghost number. The ghost number-$0$ is associated the arbitrary
invariant counterterms,
\begin{eqnarray}
\Delta^{(0)}
&=&\overbrace{(\kappa z_\kappa\frac{\partial}{\partial\kappa}
+\lambda_1 z_{\lambda_1}\frac{\partial}{\partial\lambda_1}
+\lambda_2 z_{\lambda_2}\frac{\partial}{\partial\lambda_2}
+\lambda_3 z_{\lambda_3}\frac{\partial}{\partial\lambda_3}
)\Sigma}^{\Delta^{(0)}_{\mbox{cohom}}}\\ 
&+& 
\underbrace{(z_A{\cal N}_A+z_\Psi{\cal N}_\Psi+z_\varphi{\cal N}_\varphi)
\Sigma}_{B_{\Sigma}\hat{\Delta}^{(0)}}\nonumber.
\label{solbasis}
\end{eqnarray}   
This corresponds to an expansion in the integral of the elements of the basis
\begin{equation}
\{\varphi^6, \varphi^2\bar{\Psi}\Psi, 
n_A={\cal B}_\Sigma(\hat{A}^{\natural}A),
n_\Psi=
    {\cal B}_\Sigma(\bar{\Psi}^{\natural}\Psi+\bar{\Psi}\Psi^{\natural}),
n_\varphi={\cal B}_\Sigma(\varphi^{\natural}\varphi)\}.
\label{cl_basis}
\end{equation} 
The nontrivial solutions belonging to the ghost number-$1$ sector,
$\Delta^{(1)}_{\rm cohom}$, are the possible anomalies. However, as it
was argued in \cite{CFNP} that in view of \cite{Bran, Bar}, the cohomology
of this sector is empty or at most contains Abelian ghosts, which based
on  \cite{Band}, do not contribute to gauge anomalies.
This was sufficient \cite{CFNP} to conclude the absence of
gauge anomalies and the validity of ST identities to all orders
of the vertex functional $S(\Gamma)=0$. Hence, there followed the 
renormalizability of the model.
The quantum version of (\ref{cl_trace}) can be written as bellow:  
\begin{equation}
\Theta_\mu{}^\mu(x)\cdot \Gamma \sim w^{\Phi}(x)\Gamma + 
\partial_\mu[\Lambda^\mu(x)\cdot\Gamma]+\Delta \cdot \Gamma\,, 
\label{qu_trace}
\end{equation}
where the term $\Delta\cdot\Gamma$ represents the breaking due
to radiative correction. From the absence of gauge anomalies, 
it can be expanded \cite{CFNP} by using the quantum version of basis
(\ref{cl_basis}). Once it represents the
scale breaking, its coefficients are the $\beta$-funtions  and
the anomalous dimensions, 
\begin{align}
e\,\Theta _\mu ^{~\mu}\left(x\right) \cdot \Gamma \sim& 
\overbrace{\left\{\beta_{\lambda_1}n_{\lambda_1} +
\beta_{\lambda_2}n_{\lambda_2}+\beta_{\lambda_3}n_{\lambda_3}-
\gamma_A n_A-\gamma_\psi n_\psi-\gamma_\varphi n_\varphi
 \right\} \cdot \Gamma }^{\Delta\cdot\Gamma}\nonumber \es 
&+\,w^{\Phi}
\left( x\right) \cdot \Gamma +\partial _\mu 
\,\left[ \Lambda ^\mu \left( x\right) \cdot \Gamma \right]\,, 
\label{qu_trace_basis}
\end{align}
where $n_{\lambda_k} \cdot \Gamma ~: {\int}d^3x~n_{\lambda_k} \cdot\Gamma 
={{\frac{\partial \Gamma }{\partial \lambda_k}}}$.
The absence of a $\beta_\kappa$-function corresponding to the Chern-Simons
action in the basis indicates the vanishing of the
$\beta_\kappa=0$
what proves the nonrenormalization of the Chern-Simons term.

In order to extend the proof of the conformal symmetry to the quantum level,
it is necessary to verify whether the properties of the trace 
(\ref{cl_trace})  hold for (\ref{qu_trace_basis}). Hence, it must be a 
vanishing trace or, at most, be in agreement with the form (\ref{ptrace}).
By observing the r.h.s. of the quantum
version of the trace (\ref{qu_trace_basis}), one finds that,
up to terms in dimensionful couplings and mass, the term 
$w^{(\Phi)}\left( x\right) \cdot \Gamma$ vanishes on-shell. The same
occurs for the terms in $\Delta\cdot\Gamma$ which contains the factors
in the form $\gamma_{\Psi}$, $\gamma_{A}$ and $\gamma_{\phi}$, because
they are physically trivial. There remain the $\beta$-functions terms 
together with the dimension 3 insertion $\partial_\mu[\Lambda^\mu\cdot\Gamma]$.
Once the breaking are supposed to be in the term $\Delta\cdot\Gamma$, the
divergence term must keep the same symmetries of the classical domain.
Therefore, on-shell, the breaking is due to the set of terms in the 
$\beta$-functions.
By integrating (\ref{qu_trace_basis}), we obtain the Ward identity for 
anomalous dilatation invariance, i. e., the Callan-Symanzik equation in
terms of trace identity 
\begin{align}
\displaystyle{\int}d^3x~\left[ \Theta _\mu ^{~\mu}\left(x\right) \cdot
\Gamma \right]\sim& \left(\beta_{\lambda_1} \partial_{\lambda_1}
+\beta_{\lambda_2}\partial_{\lambda_2}+\beta_{\lambda_3} 
\partial_{\lambda_3} -\gamma_A{\cal N}_A-\gamma_\psi {\cal N}_\psi 
-\gamma_\varphi {\cal N}_\varphi \right) 
\cdot \Gamma\,.
\label{CaSy}
\end{align}
In a recent paper \cite{zeda}, one imposed a dependence between the several
coupling constants of (\ref{g-inv-action1}) on the Chern-Simmons coupling 
$\kappa$~\cite{OSZ,Oe}. 
\begin{equation}
\lambda_i=\lambda_i(\kappa)\,. 
\label{condred1}
\end{equation}
This induces in the Callan-Symanzik equation (\ref{CaSy}) the reduction condition
\begin{equation}
\beta_{\lambda_i}=\beta_{\kappa}\frac{\partial\lambda_i}{\partial\kappa}\, .
\label{eqred}
\end{equation}
The dependence of (\ref{condred1}) in $\kappa$ was chosen to be polynomial. 
The dependence between the several couplings is
not rigid being corrected order by order. The various 2-loop $\beta$-functions
 \cite{AKK} were taken in (\ref{eqred}) under the requirement
(\ref{condred1}).    
Due to the vanishing of the $\beta_\kappa$, the lower order 
solution of (\ref{condred1}) are the roots $f^{(0)}_k$ of the set of polynomials 
in $\kappa$. The real solutions which keep positive 
the coefficient of $(\varphi^{\ast}_i\varphi_i)^3$ in (\ref{g-inv-action1}) 
are the physically meaningful.
The $m$-th order solution (\ref{condred1}) for the dependence of the couplings on
$\kappa$, (\ref{eqred}), can be written as
\begin{equation}
M_{kk^\prime}(f^{(0)})\chi^{(m)}_{k^\prime}
=\vartheta_k(\chi^{m-1},\chi^{m-2},\dots)\,.
\end{equation}
The matrix in the l.h.s. depends on the lower-order solution and the 
non-homogeneous term in the r.h.s. depends on the earlier solutions only. 
The existence of all orders solutions to (\ref{condred1}) is given by the
condition ${\rm det}\,M_{kk^\prime}(f^{(0)})\not= 0$,
which is shown to hold in \cite{zeda}. There followed the existence of 
regimes in the dependence of the couplings for which the whole set of 
$\beta$-functions vanish, keeping the coupling constants unrenormalized.  
 
Turning back to the equation (\ref{qu_trace}), one can see that on-shell,
in the flat space limit, the remaining
non-vanishing terms were exactly those which depend on the $\beta$-functions.
Therefore, one can see that the
conformal symmetry can be extended to the quantum domain. From the 
renormalizability
 of the model, it can be stated that the non Abelian 
Chern-Simons-matter model exhibits asymptotic 
conformal invariant solutions to the associated Callan-Symanzik equation.  

Hence, one can draw a majorconclusion.
A  suitable criterion \cite{Polc}, based on a local trace condition on the 
energy-momentum tensor, was used to show the asymptotic conformal symmetry
in the classical domain.  Such a criterion showed to be a shortcut for extend
the proof to the quantum level.
From the results of ref. \cite{CFNP}, namely, the renormalizability of the model
(\ref{g-inv-action1}), the trace of the energy-momentum tensor with the scale
breaking and the nonrenormalizabity of the Chern-Simons coupling constant, 
the framework to the proof in the quantum domain could be established. 
A quantum version of the local 
trace relation could be written which showed to have breaking due to radiative 
corrections. In a previous work, \cite{zeda}, it was shown for 
(\ref{g-inv-action1}), the existence of regimes in the dependence of the several
couplings in $\kappa$, in which the whole set of $\beta$-functions vanish 
yielding the trace relation to vanish on shell in the assymptotic limit.
The proof for cases other than $SU(2)$ would be identical; the questionable
argument is the existence of  regimes where the dependence between the couplings
lead to vanishing $\beta$-functions. However, the absence of physically 
meaningful solutions, expressed by the determinant condition, would be 
accidental. Nevertheless, it could occur for a particular value of $n$. 


I am thankful to S.D.Odintsov for the suggestion of the present analysis. 
Thanks to GFT-UCP for the friendly atmosphere and generous reception. 
I am specially greatfull to O. Piguet, J.A. Helay\"el-Neto and
D.H.T. Franco for the illuminating discussions and stimulating presence.
Thanks to the staff of the DME-PUC.



\begin{thebibliography}{99}
\bibitem{bbr-pr91} D. Birmingham, M. Blau, M. Rakowski and G. Thompson, 
{\sf Phys. Rep.~}209 (1991) 129;
\bibitem{s-btc} S. Deser, R. Jackiw and S. Templeton, {\sf Ann. of 
Phys.~}140 (1982) 372;\\ A.S. Schwarz, Baku International Topological 
Conference, Abstracts, vol.2, p.345 (1987);\\ E. Witten, {\sf Commun. Math. 
Phys.~}117 (1988) 353, {\sf Commun. Math. Phys.~}118 (1988) 601, {\sf 
Commun. Math. Phys.~}121 (1989) 351; 
\bibitem{witten-grav} E. Witten, {\sf Nucl. Phys.~}B311
(1988)46; {\sf Phys. Lett.~}B206 (1988) 601;\\ S. Deser, J. McCarthy and Z. 
Yang, {\sf Phys. Lett.~}B222 (1989) 61; 
\bibitem{guadagnini}  E. Guadagnini, M. Martellini and M. Mintchev, 
{\sf Phys. Lett.~}B227 (1989) 111, {\sf Nucl. Phys.~}B330 (1990) 575; 
\bibitem{blasi}  A. Blasi and R. Collina, {\sf Nucl. Phys.~}B345
(1990) 472;\\ F. Delduc, C. Lucchesi, O. Piguet and S.P. Sorella, {\sf Nucl. 
Phys.~}B346 (1990) 313; \\C. Lucchesi and O. Piguet, {\sf Nucl. Phys.~}B381 
(1992) 281;
\bibitem{CFNP0} O.M. Del Cima, D.H.T. Franco, J.A. Helay\"{e}l-Neto 
and O. Piguet, {\sf Lett.Math.Phys.~}47 (1999) 265; 
\bibitem{maggiore}  A. Blasi, N. Maggiore and S.P. Sorella, 
{\sf Phys. Lett.~}B285 (1992) 54;
\bibitem{CFNP} O.M. Del Cima, D.H.T. Franco, J.A. Helay\"{e}l-Neto 
and O. Piguet, {\sf JHEP~}02 (1998) 002; 
\bibitem{pisor}  O. Piguet and S.P. Sorella, ``{\sf Algebraic
Renormalization,}'' Lecture Notes in Physics, m28, Springer-Verlag, Berlin,
Heidelberg, 1995;
\bibitem{CSW} W. Chen, G.W. Semenoff and Y.S. Wu, {\sf Phys. Rev.~}D44 (1991) 
1625;
\bibitem{zeda} J.L. Acebal, D.H.T. Franco {\sf Phys. Lett. B~}502 (2001) 
316
\bibitem{Odin} S.D. Odintsov, {\sf Z.Phys. C~}54 (1992) 527;
\bibitem{Polc} J. Polchinski {\sf Nucl. Phys. B} 303 (1988) 226; 
\bibitem{Zumi} B. Zumino  {\sf 1970 Brandeis Lectures} ``
{\em Lectures on Elementary Particles and QuantumField Theory} ´´
eds. S. Deser, M. Grisaru and Pendleton, MIT Press Cambridge
\bibitem{Iorio} A. Iorio, L. O'Raifeartaigh, I. Sachs, C. Wiesendager 
{\sf Nucl. Phys. B} 495 (1997)  433;
\bibitem{CaCoJa} C. Callan, S. Coleman, R. Jackiw {\sf Ann. Phys.~}59, 
(1970) 42 
\bibitem{Gatto} S. Ferrara, R. Gatto, A.F.Grillo {\em Conformal Algebra in
Space-Time}, Springer Tracts in Modern Physics 67 (1973)
\bibitem{ZJ} J. Zinn-Justin {\em Quantum Field Theory and Critical Phenomena}
Clarendon Press - Oxford (1994)
\bibitem{Bran} F. Brandt, N. Dragon, M. Kreuzer {\sf Nucl. Phys. B} 340 
(1990) 187; 
\bibitem{Bar} G. Barnich, F. Brandt, M. Henneaux {\sf Nucl. Phys. B} 445
(1995) 357;
\bibitem{Band} G. Bandelloni, C. Becchi, A. Blasi, R. Collina 
{\sf Ann. Inst. Henri Poincar\'e} 28 (1978) 225;
\bibitem{OSZ} 
R. Oehme, K. Sibold and W. Zimmermann, {\sf Phys. Lett.~}B153 (1985) 142;
\bibitem{Oe} R. Oehme, {\sf Prog. Theor. Phys. Suppl.~}86 (1986) 215;
\bibitem{AKK} L.V. Avdeev, D.I. Kazakov, I.N. Kondrashuk, {\sf Nucl. 
Phys.~}B391 (1993) 333;
\end{thebibliography}
\end{document}